# Assessment of Quantitative Cyber-Physical Reliability of SCADA Systems in Autonomous Vehicle to Grid (V2G) Capable Smart Grids

Md Abdul Gaffar
IEEE Professional Community
Auburn Hills, USA
e-mail: mdabdul.gaffar@ieee.org

*Abstract*—The integration of electric vehicles (EVs) into power grids via Vehicle-to-Grid (V2G) system technology is increasing day by day, but these phenomena present both advantages and disadvantages. V2G can increase grid reliability by providing distributed energy storage and ancillary services. However, on the other hand, it has a scope that encompasses the cyber-physical attack surface of the national power grid, introducing new vulnerabilities in monitoring and supervisory control and data acquisition (SCADA) systems. This paper investigates the maliciousness caused by Autonomous Vehicle to Grid (AV2G) communication infrastructures and assesses their impacts on SCADA system reliability. This paper presents a quantitative reliability assessment using Bayesian attack graph [8]s combined with probabilistic capacity outage modeling based on IEEE RTS-79 system data. This work presents how AV2G-based attacks degrade system performance by using Monte Carlo simulations method, highlighting the need for cybersecurity-hardening strategies in smart grid design.

## I. Introduction

The increasing electrification of the transportation sector has redesigned the landscape of the energy systems with Artificial Intelligence, a Bi-directional power flow system such as Vehicle-to-Grid (V2G). Particularly, Automated Vehicle-to-Grid (AV2G) systems have been significantly considered due to their ability to integrate electric vehicles (EVs) as distributed energy storage sources within the broader power grid. Through bidirectional power flow, AV2G enables EVs not only to draw energy for charging but also to supply stored energy back to the grid during periods of peak demand. This model contributes in the demand based flexibility, regulation of the frequency of the power grid, voltage stability and over all enhancement of grid resilience [1]–[3].

Because of the rapid growth of AV2G technology, the possibilities of the introduction of various vulnerabilities of data-driven communication and automation also increasing because of their complex cyber-physical interdependencies. The Supervisory Control and Data Acquisition (SCADA) system is considered as the core components of these interdependencies, which coordinates signals between distributed EV chargers, substations, aggregators, and utility operators. While SCADA systems were designed for centralized and static control environments, it is now facing new challenges in ensuring energy security, real time in an increasingly decentralized and dynamic ecosystem [4]. The cyber-attack surface of the grid's the attack surface of the grid's cyber infrastructure is significantly enlarged since each EV effectively is being a network endpoint.

The primary cybersecurity threats introduced by AV2G systems have been identified and investigated in several studies. For example, man-in-the-middle attacks during handshake validation in distributed denial-of-service (DDoS) attacks. Also, hoaxed control commands, and firmware tampering in EV supply equipment (EVSE) are considered as primary cybersecurity threats in Grid's infrastructure [5][6]. Specifically, the interface between EVs and grid operators heavily depends on open communication protocols such as OCPP (Open Charge Point Protocol) and Smart Energy Profile 2.0 (SEP 2.0), which, if insufficiently encrypted or authenticated, can serve as access points for malicious phenomena [7].

Notably, Conventional perimeter-based defense mechanism of grid can be complicated due to the dynamic and mobile nature of EVs. It is difficult to authenticate EV's behavior and data streams due to their continuous change of network. The potential compromise of even a single EV or aggregator node can lead to cascading disruptions across both cyber and physical domains, as shown by simulated cyberattack scenarios in existing literature [8].

Recent studies have focused on developing thorough models to predict and quantify the impacts of cyberattacks on power system operation. Recent research has offered a probabilistic framework for understanding the growing recognition of this type of vulnerability, focusing on developing a comprehensive model using a Bayesian attack graph to estimate and quantify the impacts of cyberattacks on power system operations and propagate through network topologies [9]. These graphs model attack paths and quantify the likelihood of successful exploitation, aiding in both risk assessment and prioritization of defense strategies.

Furthermore, hybrid simulation platforms that couple power system dynamics with network communication protocols have been employed to assess the cascading effects of cyberattacks. Tools like Real-Time Digital Simulators (RTDS), NS-3, and MATLAB/Simulink-based co-simulations enable researchers to test various threat vectors, such as false data injection (FDI) and load manipulation attacks, under realistic grid operating conditions [10]. These models reveal that even low-probability cyber events can result in substantial physical consequences, including frequency excursions, voltage collapse, and partial grid outages.

To counter these threats, several resilience-enhancing mechanisms have been proposed. These include anomaly-based intrusion detection systems (IDS) tailored for V2G traffic patterns, decentralized control algorithms that limit the effect of compromised nodes, and secure firmware-over-the-air (FOTA) updates for EVSEs [11]–[13]. Additionally, blockchain-based transaction verification has gained interest as a means to secure data exchange between EVs and grid aggregators without relying on a central trusted authority.

Another promising approach involves the application of dynamic islanding techniques, where segments of the grid can autonomously isolate themselves upon detection of suspicious activity, thereby limiting the spread of cyber intrusions. Dynamic authentication schemes based on behavioral fingerprints or AI-driven network segmentation also show potential for enhancing the cybersecurity posture of AV2G environments [14].

Building upon this rich foundation of prior work, the present research aims to systematically assess the reliability degradation of SCADA systems under AV2G-specific cyber threats. By incorporating Bayesian network models with real-world EV load data and SCADA communication patterns, the study provides a quantitative framework for estimating the impact of cyber intrusions on key grid performance indicators such as frequency stability, unserved energy, and system reliability indices (e.g., SAIDI/SAIFI). This approach not only enhances the theoretical understanding of cyber-physical interdependencies but also offers practical insights for grid operators in developing risk aware V2G deployment strategies.

## II. BACKGROUND AND METHODOLOGY

SCADA systems are considered as the most crucial components of power system automation and control and power system reliability. Their exposure to cyber threats is increasing since the integration of V2G system technology due to the extended attack surface. Bayesian attack graph [15] have been developed to emerge as a tool for modeling multi-stage cyber threats, representing dependencies between different events. In our study, power generation reliability models such as Capacity Outage Probability Tables (COPT) and Loss of Load Probability (LOLP) are used to assess the adequacy of generation under different operating conditions.

This study combines Bayesian network modeling of cyberattack propagation with a probabilistic generation adequacy assessment using the Capacity Outage Probability Table (COPT) approach for the To quantitative assessment of the impact of AV2G-related cyber threats on power system reliability. The methodology involves two interlinked components:

a) The calculation of cyberattack propagation likelihood through a Bayesian attack graph,
b) The computation of system LOLE under varying cyber threat scenarios.

Acyclic graph is used to represent the attack surface for AV2G-to-grid where each node corresponds to a cyber event (e.g., EV Charger breach, Aggregator compromise etc.). $A_1, A_2, ..., A_n$ are assumed as the sequence of attack stages. Chain rule of conditional probability is considered for the computation of the joint probability of successful attack propagation:

$$P(A_1, A_2, ..., A_n) = P(A_1) \times P(A_2|A_1) \times P(A_3|A_2) \times ... \times P(A_n|A_{n-1}) \quad (1)$$

For example, with nodes:
- $A_1$: EV Charger Compromise
- $A_2$: Aggregator Breach
- $A_3$: SCADA Intrusion
- $A_4$: Grid Disruption

The computation of the attack propagation will be as follows:

$$P\_grid\ disruption = P(A_1) \times P(A_2|A_1) \times P(A_3|A_2) \times P(A_4|A_3) \quad (2)$$

Known vulnerability scores are used to derive each conditional probability (e.g., CVSS metrics) or are estimated from literature and simulations [15].

The impact of cyber-induced outages into the COPT are incorporated, by which all possible generation outage states and their associated probabilities are listed. The LOLE is computed by summing the probability of all states where available capacity falls short of the forecasted load:

$$LOLE = \sum P_i \times D_i \quad (3)$$

Where:
- $P_i$ is the probability of outage state i
- $D_i$ is the duration (in days/year) that load exceeds available generation in state i

A cyber-induced de-rating factor $\delta$ is applied to certain states to account for cyber threats. For instance, if a successful cyberattack restricts x% of generation units, the outage state probabilities and their severities are adjusted as follows:

$$P_i\_cyber = P_i \times (1 + \delta)$$
$$C_i\_available = C_i \times (1 - \delta) \quad (4)$$

Where:
- $\delta \in [0,1]$ is the cyber compromise factor (e.g., 0.05 for 5%)
- $C_i\_available$ is the reduced available capacity in state i

This approach provides a novel cyber-physical approach to reliability assessment by combining conditional probabilities form Bayesian attack graphs into generation adequacy modeling. This framework can be dynamically adjusted as new vulnerabilities or mitigation mechanisms emerge, making it extensible for real-world AV2G deployment.

This paper describes a comprehensive, simulation-based analysis of how Autonomous Vehicle-to-Grid (AV2G) systems can introduce new cyber-physical vulnerabilities into smart grid environments, with a specific focus on Supervisory Control and Data Acquisition (SCADA) systems. This work has blended probabilistic modeling via Bayesian attack graphs and conventional reliability tools such as Capacity Outage Probability Tables (COPT) and Monte Carlo simulations to quantify the swamping effects of cyber threats on power system reliability. A detailed breakdown of attack propagation routes, negative effect on the power system reliability, and architectural recommendations for cybersecurity-hardening of AV2G-integrated smart grids can be considered as key contributions in this study.

In this study, a layered methodology was developed that combines attack graph modeling with generation adequacy analysis to investigate the propagation of cyber attacks causing compromise of grid functions. This combined methodology allows for both qualitative and quantitative exploration of the risks posed by AV2G-enabled attack surfaces.

The structure of the Bayesian attack graph which is used in the study was summarized [15]. Each node resembles to a critical compromise point in the AV2G–SCADA communication chain, and the conditional probabilities reflect published vulnerability indices and cyber threat likelihoods.

Table 1: Bayesian Attack Graph Node Definitions and Probabilities

| Node | Description | Conditional Probability |
|---|---|---|
| EV Charger | Initial intrusion through V2G charger firmware | 0.07 |
| Aggregator | Attack on centralized control hub | 0.04 (given EV compromise) |
| SCADA | Compromised aggregator sends malicious SCADA commands | 0.06 (given aggregator breach) |
| Grid Disruption | Final failure in dispatch/load balancing | 0.08 (given SCADA intrusion) |

The simulation setup for the reliability evaluation is listed in Table 2. Parameters such as the total number of generators, load horizon, attack timing, and generator availability rates are considered for the simulation. These settings support IEEE RTS-79 standards and have been modified to simulate cyber-influenced generator outages and reduced operational capacity. Such documentation validates reproducibility and reinforces the relevance of the scenario modeling.

Table 2: Simulation Parameters and Assumptions

| Parameter | Value | Description |
|---|---|---|
| Total Generators | 11 | Based on IEEE RTS-79 Test System |
| Annual Simulation Hours | 8760 | Full-year simulation resolution |
| Base LOLP (No Attack) | <0.01 | System reliability baseline |
| Attack Window | Hour 4380 | Simulated cyber attack peak timing |
| Generator Availability | 0.95 avg | Nominal online probability without attack |
| Cyber-Affected Availability | 0.88 avg | Degraded availability during attack |

III. RESULT AND DISCUSSION

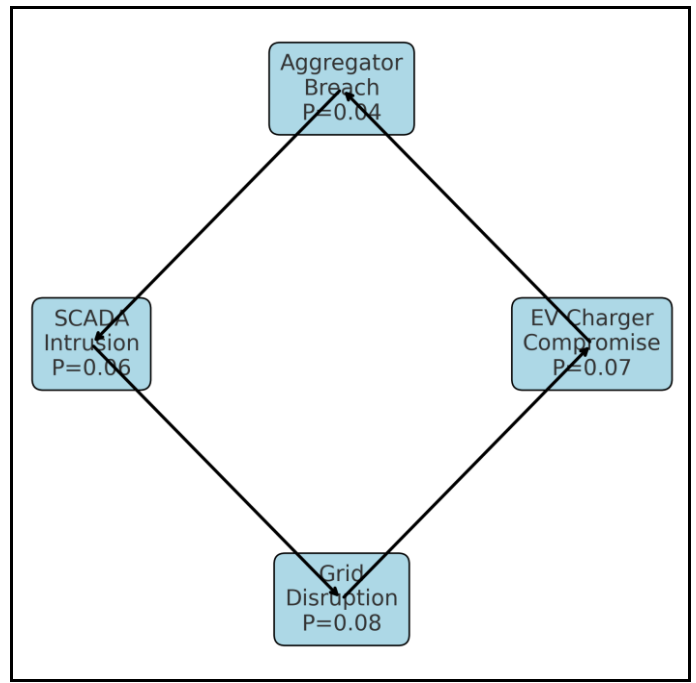

**Fig. 1** Bayesian Attack Graph for AV2G System

Each node in the Bayesian Attack graph in Fig.1 represented a stage in the attack chain and was assigned a conditional probability based on real-world vulnerability metrics. To evaluate system reliability under these threats, A COPT is constructed using generator availability and capacity data from IEEE RTS-79 to assess system reliability.

This graph was used to map the multi-stage intrusion path from the EV charger to the SCADA control layer. This model includes conditional probabilities for each stage (Table 1) - EV charger compromise, aggregator breach, SCADA intrusion, and grid disruption based on known vulnerability metrics. The graph supports risk assessment and prioritization of countermeasures by probabilistically linking these nodes. The model emphasized that even a seemingly minor EV entry point can lead to significant grid-level disruptions if not properly monitored and assessed.

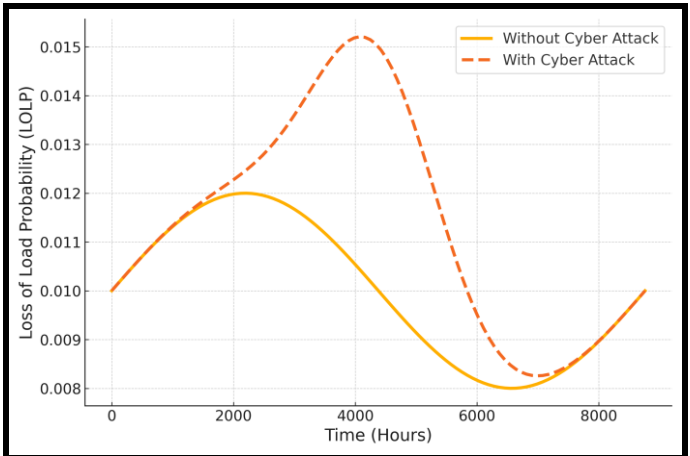

**Fig. 2** Loss of Load Probability (LOLP) Over time

This figure (Fig. 2) illustrates the increase in Loss of Load Probability (LOLP) during a simulated cyber-attack on SCADA systems […] integrated with AV2G. While with the integration of AV2G within the grid, the peak disruption occurs around hour 4380, lining up with expected summer peak demand. This information suggests that in the peak hour or peak load seasons, the time-of-year should be carefully accounted to avoid the margin violation by the attack leading to cascading outage. In this figure, it was clearly shown even a small probability of compromise can increase the LOLP by around 2 times, revealing the breakability of system-level reliability when cyber threats coincide with demand surges.

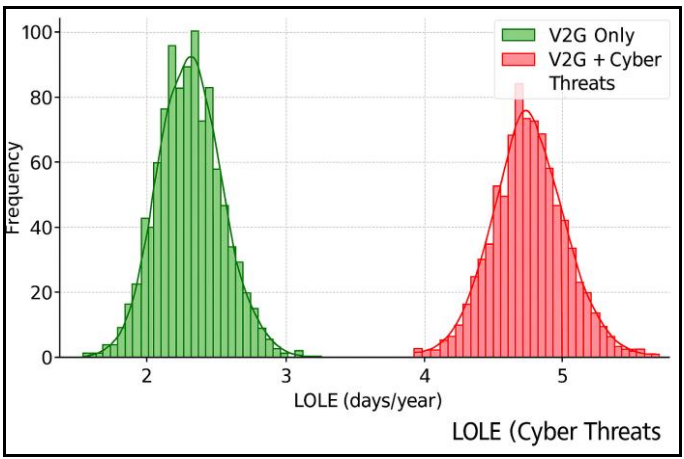

**Fig. 3**. Monte Carlo Simulation of LOLE Outcomes

The Monte Carlo simulations were presented in Fig. 3. to present a probabilistic analysis. The distribution of LOLE values under only V2G conditions appeared firmly around 2.3 days/year, while the cyber-threats compromised case shows a wider and right-shifted distribution, centering near 4.8 days/year. This implies that cyber threats because of V2G not only worsen reliability but also introduce greater uncertainty in grid performance.

Additionally, this study compared the LOLE under three grid configurations: baseline (3.5 days/year), secure V2G (2.3 days/year), and V2G with cyber threats (4.8 days/year), as shown in Fig. 4. While V2G improves reliability when secure, the presence of even modest cyber threats reverses this benefit, increasing outage risks. The result underscores the need for cybersecurity as a precondition for realizing V2G reliability gains.

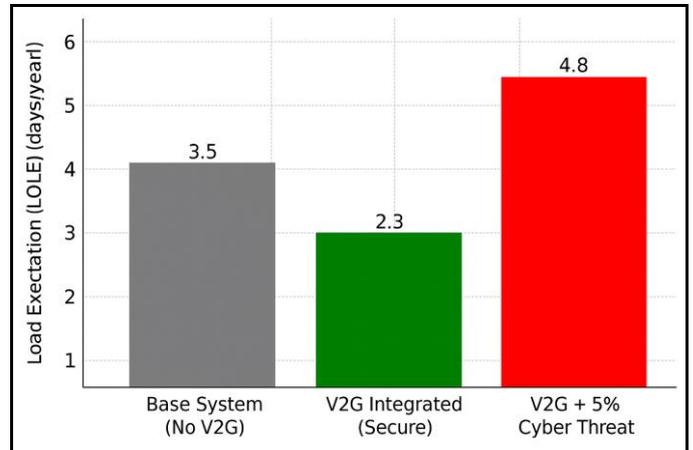

**Fig. 4**. Comparative Capacity Outage Probability Table (COPT)

Based on the theoretical foundations of Section 2, system behavior under cyber threat conditions by using MATLAB-based models was simulated for the computation of unavailable generator and the negative impact of cyber-influenced outage probabilities on system-level reliability metrics such as Loss of Load Expectation (LOLE) and LOLP. In can be clearly noticed from the simulation results that even moderate cyber-attack probabilities lead to significant risk elevation in the power system. These findings highlight the imperative need for cybersecurity-developed system architecture as detailed in Figure 6, and confirm the critical role of real-time intrusion detection and segmented communication channels in AV2G systems.

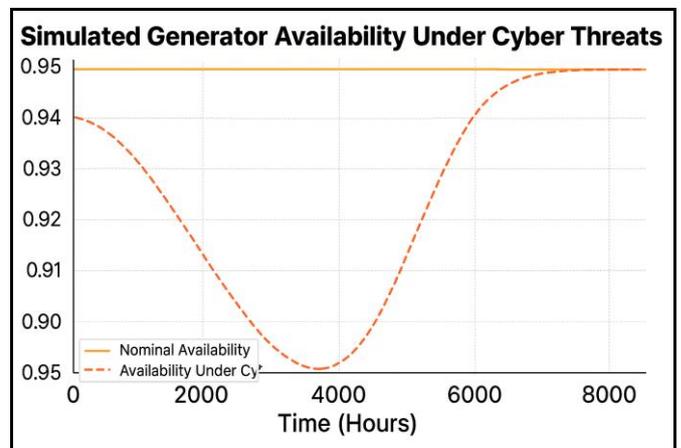

**Fig. 5** Generator Availability Under Cyber Attack

This simulation results in Fig. 5 shows how generator availability decreases when SCADA control paths are compromised due to a cyber intrusion originating from an

AV2G communication breach. The decline in availability highlights how control path interruptions were highlighted by the decline in availability of generator-such as deceived dispatch commands or deferred feedback from sensors—can simulate mechanical failure without any physical fault. This cyber-induced de-rating creates artificial extra stress on standby capacity. Overall system margin shrinks with the report of more "offline" generators due to the misinformation; hence causing the risk of load shedding increases. This figure emphasizes the importance of separate back-up critical generator communication channels from V2G-accessible layers.

This Simulation showed that the reliability benefits of AV2G can be substantially eroded by cyber threats. Without attacks, LOLP remained below 0.01 for most of the year without any cyber-attack, while LOLP spiked to 0.02–0.025 under a mid-year cyber-attack. Generator availability also decreased by 5–8%, significantly affecting grid resiliency.

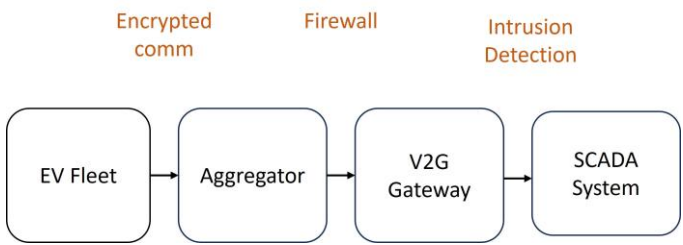

**Fig. 6** Cybersecurity-Enhanced AV2G-SCADA Architecture

Figure 6 illustrates a secure control architecture that connects EVs to the grid via V2G interfaces. This framework would comprise three protective layers: encrypted communication, firewalls, and intrusion detection. These secure defenses ensure that commands and data exchanged between On-Board Chargers (OBCs) and SCADA remain trustworthy, thereby preserving grid reliability and mitigating the cyber vulnerabilities inherent to V2G deployment. The integrity of data and commands can be protected as they flow through the aggregator and V2G gateway, thereby safeguarding the reliability gains. Together, these results corroborate the critical need for embedding robust cybersecurity into the control strategies of On-Board Chargers (OBCs) to ensure that V2G technologies can fulfill their promise as a grid reliability asset rather than a vulnerability.

## IV. CONCLUSION

The study demonstrates that the cyber intrusion caused by AV2G, if not lessened, can interrupt otherwise valuable energy services. Encrypted communication, real-time anomaly detection, and resilient SCADA architecture are vital in the power system. Probabilistic modeling tools, such as Bayesian graphs, offered meaningful ways to quantify risk and guide grid modernization efforts, integrating cyber threat analysis with power system reliability simulations. The use of Bayesian attack graphs, COPT, and Monte Carlo methods provides a robust framework for evaluating AV2G impact on SCADA systems, indicating the urgent need for proactive cybersecurity integration to ensure the reliability of future smart grids.